\documentclass[journal]{IEEEtran}
\ifCLASSINFOpdf
  \usepackage[pdftex]{graphicx}
  \graphicspath{{figure/}}
\else
\fi
\usepackage{amsmath}
\usepackage{mathptmx}
\usepackage{mathtools}
\allowdisplaybreaks

\begin{document}
%
\title{Coordinated Primary Frequency Regulation and Grid-Forming Control for Wind Turbine Generators}
%
%
%

\author{Meng~Chen,~\IEEEmembership{Member,~IEEE,}
        Yufei~Xi,~\IEEEmembership{Member,~IEEE,}
        Lin~Cheng,~\IEEEmembership{Member,~IEEE,}
        Florian~D\"{o}rfler,~\IEEEmembership{Senior~Member,~IEEE,}~and~Ioannis~Lestas,~\IEEEmembership{Member,~IEEE}
\thanks{Meng Chen and Ioannis Lestas are with the Department of Engineering, University of Cambridge, Cambridge CB2 1PZ, United Kingdom 

Yufei Xi and Lin Cheng are with the State Key Laboratory of Power System Operation and Control (Department of Electrical Engineering, Tsinghua University), Beijing 100084, China 

Florian D\"{o}rfler is with the Automatic Control Laboratory, ETH Z\"{u}rich, 8092 Z\"{u}rich, Switzerland 
}}

%
%

\markboth{}%
{Shell \MakeLowercase{\textit{et al.}}: Bare Demo of IEEEtran.cls for IEEE Journals}
%



\maketitle

\begin{abstract}
Conventional grid-forming (GFM) control strategies often treat the DC source as an unconstrained link, creating mismatches when applied to the wind turbine generators (WTGs). Focusing on primary frequency regulation, this paper systematically investigates the mismatch between the GFM-WTGs behavior and the droop-based primary frequency regulation. To address this issue, a novel coordination strategy between WTG primary frequency regulation and GFM control is proposed. By establishing well-designed relationships among the power-tracking coefficient, power set-point, and frequency deviation, the proposed strategy enables GFM-WTGs to participate consistently in primary frequency regulation within predefined frequency limits while maintaining appropriate power points and effectively utilizing the allowable power reserve. Furthermore, the proposed method preserves the control structure and dynamic performance of conventional GFM control and inherently adapts to varying wind-speed conditions. Comparative case studies under different operating scenarios demonstrate the effectiveness and superiority of the proposed strategy.
\end{abstract}

\begin{IEEEkeywords}
Wind turbine, grid-forming control, primary frequency regulation, droop characteristics
\end{IEEEkeywords}

%
\IEEEpeerreviewmaketitle

\section{Introduction}

\IEEEPARstart{T}{he} increasing penetration of converter-interfaced renewable generation is fundamentally changing the dynamic characteristics of modern power systems \cite{Chen2020,Blaabjerg2023}. As conventional synchronous generators (SGs) are gradually displaced by wind and solar resources, the aggregate system frequency support capability is significantly reduced \cite{Wang2018,Gonzalez2023}. As a result, modern power grids face increasing frequency stability risks following power imbalances. To mitigate these issues, renewable generation units are expected to actively participate in ancillary services, particularly primary frequency regulation, in addition to energy production.

Wind turbine generators (WTGs) possess substantial kinetic energy storage in their rotating masses and thus have the potential to contribute to system frequency regulation \cite{Blaabjerg2024}. In conventional grid-following (GFL) WTGs, frequency support is typically achieved by modifying the active power reference away from the maximum power point tracking (MPPT) operating point \cite{Wang2018}. Existing approaches can be broadly classified into trajectory-based \cite{Yang2018} and feedback-based methods \cite{Bonfiglio2019}. Trajectory-based methods inject predefined power during disturbances, but often suffer from rigid recovery dynamics, leading to secondary frequency dips during rotor speed restoration \cite{Zhou2023,Heidari2025}. Feedback-based methods improve adaptability by linking active power to instantaneous frequency deviations \cite{Tang2024}. Nonetheless, following a sustained frequency decrease, a WTG operating in MPPT mode tends to settle at an unexpectedly lower rotor speed \cite{Bao2023}. To address this issue, WTGs can be operated in a deloaded mode to maintain a continuous power reserve for primary frequency regulation \cite{Zhao2025a}. Nevertheless, the dependence on phase-locked loops for synchronization limits their performance in weak grids and large-scale wind energy integration scenarios \cite{Yang2024,Zhao2025}.

Recently, grid-forming (GFM) control has emerged as a promising solution for future power systems with high renewable penetration. By establishing voltage and frequency autonomously, GFM converters can emulate SG behavior and provide frequency support without relying on a stiff grid reference \cite{Chen2024}. Numerous GFM strategies have been proposed to enhance inertia emulation \cite{Jiang2025}, transient stability \cite{Wang2023}, and frequency regulation performance \cite{Alghamdi2021}. Despite their differences in control implementation, most GFM schemes exhibit explicit droop-based active power-frequency characteristics in primary frequency regulation. However, these GFM schemes usually assume an ideal DC source, neglecting the aerodynamic constraints of WTGs. This leads to mismatched regulation when directly applied to WTG systems.

Although this mismatch can be partially mitigated by incorporating WTG aerodynamics into GFM power references similar to GFL-WTG schemes \cite{Chen2022,Liu2024}, fundamental challenges remain. First, the droop-imposed operating point may conflict with rotor-speed-dependent power characteristics \cite{Gonzalez2024}, thus leading to harmful responses where WTGs reduce power during frequency drops \cite{Zhang2025}. Second, although deloading can provide power reserve, existing strategies generally lack explicit coordination between frequency regulation requirements and the allowable aerodynamic operating range \cite{Lyu2024}. Consequently, the reserved capability may be either underutilized or exceeded during upward/downward frequency regulation. Third, since WTG operating characteristics vary continuously with wind speed, existing strategies often exhibit limited adaptability across different wind speeds \cite{Yuan2024}.

These observations indicate that a key challenge lies in the coordinated integration of frequency regulation requirements, GFM control, and WTG aerodynamic characteristics. Motivated by this, this paper proposes a coordinated control framework for GFM permanent magnet synchronous generator (PMSG)-based WTG systems. The proposed method establishes an explicit mapping between grid frequency, WTG operating state, and active power reference through a frequency-dependent power-tracking mechanism.

The contributions of this paper are summarized as follows:
\begin{enumerate}
    \item The mismatch among GFM control, primary frequency regulation, and WTG characteristics is studied in detail.
    \item A coordinated active power reference design methodology is developed for GFM-PMSG-WTGs, which enables the WTGs to provide the desired primary frequency response while remaining within predefined operating limits, ensuring effective utilization of power capability and automatic power sharing among paralleled units.
    \item The proposed strategy exhibits strong adaptability to varying wind conditions and can preserve the original GFM control architecture, thereby improving the implementation practicality. 
\end{enumerate}

The remainder of this paper is organized as follows. Section \ref{section_gfm} introduces the GFM-PMSG-WTG system model. Section \ref{section_mppt} and Section \ref{section_deloading} elaborate on the mismatching issues of the typical MPPT strategy and deloading strategy, respectively. The proposed coordinated control strategy is presented in Section \ref{section_proposed}. Section \ref{section_case} validates the proposed method through case studies under various operating conditions. Finally, Section \ref{section_conclusion} concludes the paper.

\section{Grid-Forming PMSG-Based Wind Turbine Generator System}\label{section_gfm}

\begin{figure*}[!t]
\centering
\includegraphics[width=\textwidth]{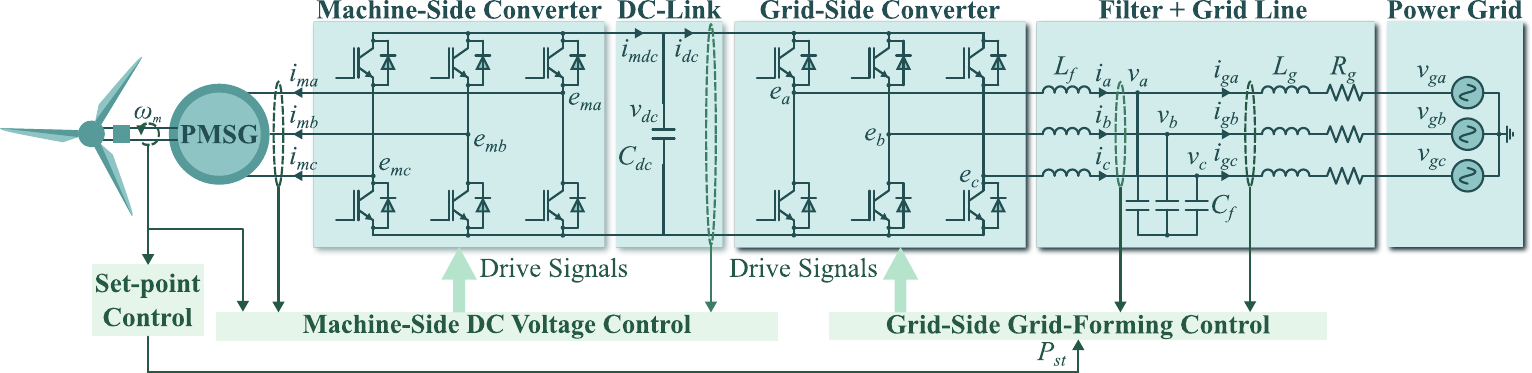}
\caption{Configuration of a grid-forming permanent magnet synchronous generator-based wind turbine generator system .}
\label{fig_StudiedSystem}
\end{figure*}

\begin{figure}[!t]
\centering
\includegraphics[width=\columnwidth]{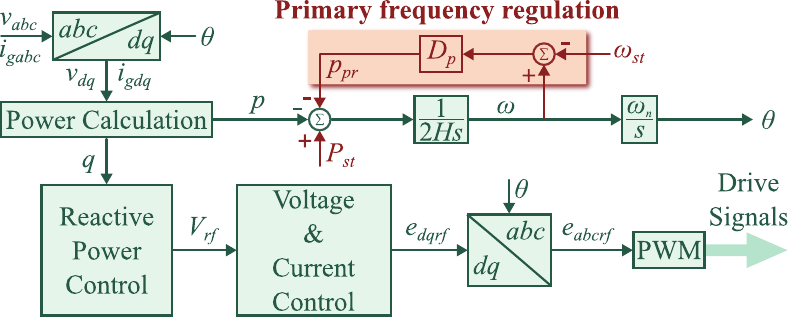}
\caption{Control block diagram of grid-side grid-forming control.}
\label{fig_Control}
\end{figure}

The topology of the GFM-PMSG-WTG system is shown in Fig. \ref{fig_StudiedSystem}. The machine-side converter (MSC) regulates the DC-link voltage $ v_{dc}$. The grid-side converter (GSC) is operated under a GFM control framework of Fig. \ref{fig_Control}. 

The power captured by the WT is expressed in p.u. form as
\begin{align}
\label{eq_pw}
    p_{w} = (1/2)\rho\pi R^2C_p(\lambda,~\beta)v_w^3/S_n
\end{align}
where $\rho$ is air density, $R$ is blade radius, $v_w$ is wind speed, and $S_n$ is the rated power. The performance coefficient $C_p$ depends on the tip-speed ratio $\lambda$ and pitch angle $\beta$, and is modeled as
\begin{align}
\label{eq_cp}
    &C_p(\lambda,~\beta) = c_1\left(c_2/\lambda_i - c_3\beta - c_4\beta^x - c_5\right)e^{-c_6/\lambda_i}\\
    &1/\lambda_i = 1/(\lambda + 0.08\beta) - 0.035/(\beta^3 + 1)
\end{align}
with empirical constants $c_1$-$c_6$ and $x$. $\lambda$ is defined as
\begin{align}
    \lambda = \omega_{tn}\omega_mR/v_w
\end{align}
where $\omega_m$ is the p.u. rotor speed and $\omega_{tn}$ is the nominal mechanical angular frequency base.

The rotor electromechanical dynamics are described by:
\begin{align}
\label{eq_domegam}
    2H_m\dot\omega_m = p_w - p 
\end{align}
where $H_m$ is the inertia constant of the combined mechanical parts and $p$ is the electrical active power output.

\begin{figure}[!t]
\centering
\includegraphics[width=0.63\columnwidth]{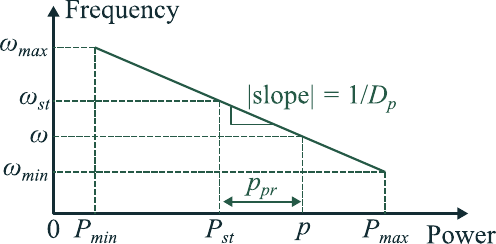}
\caption{Droop-based primary frequency regulation characteristics.}
\label{fig_Droop}
\end{figure}

The frequency of the GFM controller $\omega$ is governed by:
\begin{align}
\label{eq_domega}
    2H\dot\omega = P_{st} - p - p_{pr}
\end{align}
where $H$ is the virtual inertia constant, $P_{st}$ is the active power setpoint, $p_{pr}$ is the primary frequency regulation power, and $\omega_n$ is the nominal angular frequency. $p_{pr}$ is given by:
\begin{align}
\label{eq_ppr}
    p_{pr} = D_p(\omega - \omega_{st})
\end{align}
where $D_p$ is the droop coefficient and $\omega_{st}$ is the nominal frequency in p.u. As shown in Fig. \ref{fig_Droop}, the resulting regulation characteristic can equivalently be interpreted through reference-based formulation using ($P_{st}$, $\omega_{st}$) together with $D_p$, or limit-based formulation using ($P_{min}$, $\omega_{max}$) and ($P_{max}$, $\omega_{min}$).

Unlike energy storage systems, where $P_{st}$ can be freely dispatched, in WTGs it is constrained by aerodynamic characteristics. Therefore, $P_{st}$ becomes a function of rotor speed:
\begin{align}
\label{eq_pst}
    P_{st} = f(\omega_m)
\end{align}
which highlights the intrinsic coupling between mechanical dynamics and GFM control. As a result, frequency regulation behavior is jointly determined by aerodynamic characteristics and GFM control law. Without coordinated design of $f(\omega_m)$, turbine aerodynamics and GFM control, the operating point may deviate from the intended primary frequency response, degrading regulation performance.

\section{Limitations of MPPT-Based Grid-Forming Wind Turbine Generator Frequency Regulation}\label{section_mppt}

In existing GFM-WTG applications, one of the typical values of $P_{st}$ is selected according to the MPPT principle. Under this strategy, $P_{st}$ is directly determined by $\omega_m$ as
\begin{align}
\label{eq_fomegam_mppt}
    f(\omega_m) = p_{mppt} = K_{opt}\omega_m^3
\end{align}
where $K_{opt}$ is the optimal power-tracking coefficient determined by the aerodynamic characteristics of the WTG as
\begin{align}
    K_{opt} \coloneq 1/2\rho\pi R^5C_{popt}(\lambda_{opt},~\beta)\omega_{tn}^3/(\lambda_{opt}^3S_n).
\end{align}
Here the subscript "$opt$" denotes the optimal value. 

Combining (\ref{eq_domegam})-(\ref{eq_fomegam_mppt}), the droop-based primary frequency regulation satisfies
\begin{align}
    K_{opt}\omega_m^3 - p_w(\omega_m,~v_w) - D_p(\omega - \omega_{st}) = 0.
\end{align}
It highlights that the actual regulation characteristics are constrained by both the droop characteristic of the GFM control and the aerodynamic behavior of the wind turbine.

\begin{figure}[!t]
\centering
\includegraphics[width=\columnwidth]{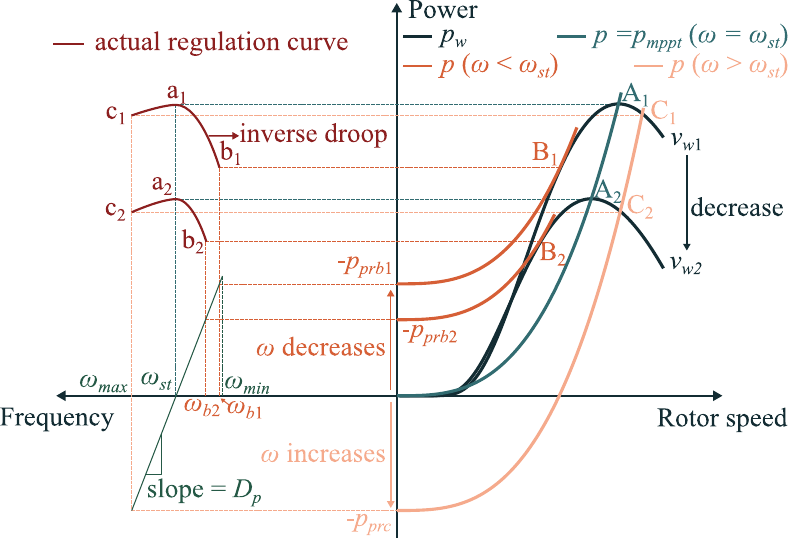}
\caption{Schematic of primary frequency regulation with MPPT control.}
\label{fig_MPPTDroop}
\end{figure}

When it always remains that $\omega = \omega_{st}$, the droop term $p_{pr}$ vanishes and the WTG operates at the MPPT point. The corresponding operating point is denoted as A$_1$ in Fig. \ref{fig_MPPTDroop} for the wind speed $v_{w1}$. Under this condition, the WTG achieves maximum energy capture and no coordination issue exists.

However, a mismatch arises during frequency deviations. Suppose a frequency disturbance that causes $\omega$ to decrease to $\omega_{b1} < \omega_{st}$. The droop characteristic increases the active power output by $-p_{prb1} = D_p(\omega_{st} - \omega_{b1})$. As a result, the new operating point is established at a lower $\omega_m$ and a lower $p$, as illustrated by the transition from A$_1$ to B$_1$ in Fig. \ref{fig_MPPTDroop}. Consequently, the actual regulation characteristic exhibits an inverse droop phenomenon as the trajectory from a$_1$ to b$_1$. This inverse droop trend fundamentally violates primary frequency regulation requirements. Instead of sustaining an increased power injection to support the network, the MPPT-GFM-WTG curtails its power contribution during under-frequency contingencies. If $\omega$ drops further below this threshold, i.e., $\omega \in [\omega_{min},~\omega_{b1})$), no feasible equilibrium exists. The rotor enters an uncontrollable deceleration, which forces an instability.

When $v_w$ decreases to $v_{w2}$, the power curve $p_w$ compresses vertically as shown in Fig. \ref{fig_MPPTDroop}. As a result, under lower wind profiles, the frequency deviation margin required to trigger the instability is severely reduced with $\omega_{b2} > \omega_{b1}$. The WTG becomes exceptionally vulnerable, demonstrating that the primary frequency regulation capability is highly volatile and tightly coupled to time-varying environmental conditions.

Conversely, during $\omega > \omega_{st}$, the droop loop commands a power reduction by $p_{prc}$, accelerating the rotor toward an over-speeding state (A$_1\rightarrow$ C$_1$ and a$_1\rightarrow$ c$_1$ for $v_{w1}$ in Fig. \ref{fig_MPPTDroop}). While this direction aligns with primary frequency regulation rules, the final limit point $C_1$ is entirely unregulated by any explicit constraint. The achieved power reduction may be either insufficient or excessive relative to the predefined range. 

\section{Limitations of Existing Deloaded GFM Wind Turbine Generator Frequency Regulation}\label{section_deloading}

\begin{figure}[!t]
\centering
\includegraphics[width=\columnwidth]{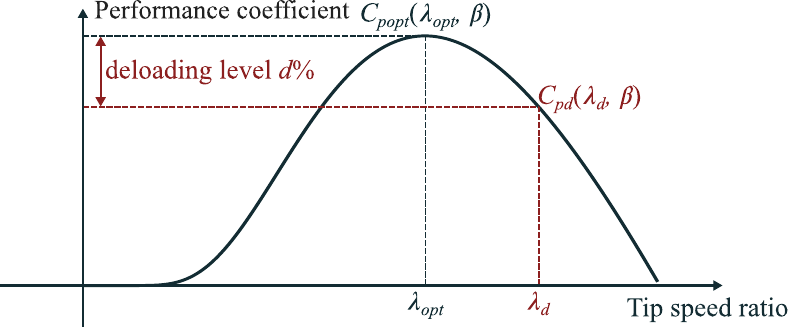}
\caption{Characteristics of performance coefficient and deloading operation.}
\label{fig_PerformanceCoefficient}
\end{figure}

To mitigate the inverse droop characteristic and subsequent instability, WTGs are proposed to operate in a deloaded mode \cite{Lyu2024}. By intentionally reducing the power output below the maximum available wind power during normal operation, a power reserve can be continuously maintained and subsequently released during frequency disturbances. Accordingly, $p_{mppt}$ is replaced with a deloaded power $p_d$ for $P_{st}$, i.e.,
\begin{align}
\label{eq_fomegam_deload}
    f(\omega_m) = p_d = K_d\omega_m^3
\end{align}
where $K_d$ represents the deloaded power-tracking coefficient. The value of $K_d$ is selected such that the WTG operates at a reduced power level under $\omega_{st}$. This operating point is characterized by a predefined deloading level $d\%$, which represents the percentage of power curtailed from $p_{mppt}$. The corresponding value of $K_d$ can be determined as
\begin{align}
    K_d \coloneq K_{opt}\left(\lambda_{opt}/\lambda_d\right)^3(1-d\%)
\end{align}
where $\lambda_d$ denotes the tip-speed ratio associated with the selected over-speed deloaded operating point. Fig. \ref{fig_PerformanceCoefficient} shows the relationship between $C_p$, $\lambda$, and $d\%$. As the operating point moves away from $\lambda_{opt}$, a power reserve required for primary frequency regulation is created.

However, the selection of an appropriate $d\%$ remains a challenging issue. Existing studies typically determine $d\%$ through empirical tuning, which lacks an explicit analytical mapping to the resulting active power-frequency regulation characteristics. As a result, the frequency regulation characteristics may still exhibit undesirable performance. To better illustrate these limitations, the following subsections investigate the regulation characteristics associated with low-deloading and high-deloading operating conditions.

\subsection{Low Deloading Level}\label{subsection_LowDeloading}

\begin{figure}[!t]
\centering
\includegraphics[width=\columnwidth]{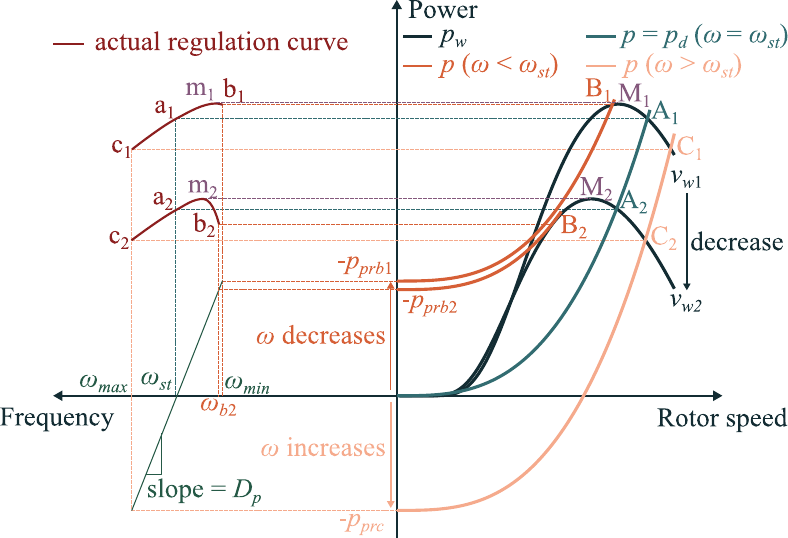}
\caption{Schematic of primary frequency regulation characteristics with a low deloading level.}
\label{fig_DeloadingLowDroop}
\end{figure}

When a relatively low deloading level $d\%$ is selected, the nominal operating point is intentionally shifted only slightly away from the maximum power point. At $\omega = \omega_{st}$, the droop term $p_{pr}$ is zero and the WTG operates at the deloaded point A$_1$ under $v_w = v_{w1}$, as shown in Fig. \ref{fig_DeloadingLowDroop}. Under this condition, the turbine curtails a minimal amount of available wind power, maintaining high energy capture efficiency but leaving a narrow upward regulation margin.

When $\omega$ decreases, the droop characteristic increases the converter active power output and the operating point moves from A$_1$ toward the MPPT point M$_1$, as shown in Fig. \ref{fig_DeloadingLowDroop}. Within this region, the deloaded operation successfully compensates for the reduction in $\omega_m$, enabling the WTG to provide a sustained active power increase as $\omega$ decreases. The frequency regulation capability is thus successfully improved by the expected droop characteristic along the path from a$_1$ to m$_1$ in Fig. \ref{fig_DeloadingLowDroop}. However, because $d\%$ is small, once the operating point reaches M$_1$, further reduction in $\omega$ forces the operating point to move into the underspeed region from M$_1$ toward B$_1$. In this region, the same inverse droop regulation trend identified in Section \ref{section_mppt} emerges again, as shown by the trajectory from m$_1$ to b$_1$ in Fig. \ref{fig_DeloadingLowDroop}.

This limitation becomes more pronounced under varying wind speed conditions. As $v_w$ decreases from $v_{w1}$ to $v_{w2}$, the stable operating region becomes highly restricted as shown in Fig. \ref{fig_DeloadingLowDroop}. The operating point can easily go beyond the limiting point B$_2$ under $\omega = \omega_{b2}$, potentially leading to excessive rotor deceleration and loss of stable operation. Therefore, low-level deloading cannot guarantee satisfactory frequency regulation performance across a wide range of wind conditions. Furthermore, during over-frequency conditions $\omega > \omega_{st}$, the required power reduction drives the operating point toward C$_1$ and C$_2$ with higher $\omega_m$ and deeper power curtailment, making the system highly susceptible to exceeding desirable power and speed operating boundaries in response to an identical $-p_{prc}$ compared with MPPT-GFM control.

\subsection{High Deloading Level}

\begin{figure}[!t]
\centering
\includegraphics[width=\columnwidth]{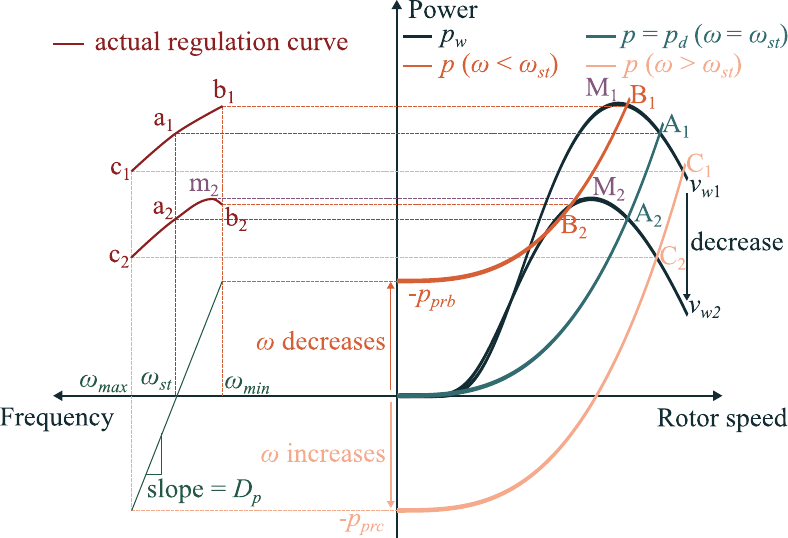}
\caption{Schematic of primary frequency regulation characteristics with a high deloading level.}
\label{fig_DeloadingHighDroop}
\end{figure}

To further eliminate the inverse droop phenomenon and avoid instability, a relatively high $d\%$ can be implemented as shown in Fig. \ref{fig_DeloadingHighDroop}. By increasing $d\%$, the nominal operating point A$_1$ under $v_{w1}$ is positioned farther down the right-side of the MPPT point M$_1$. This configuration establishes a substantially larger upward regulation margin at the expense of increased energy curtailment during normal operation.

When $\omega$ decreases, the operating point moves from A$_1$ toward M$_1$. Because the available power reserve is sufficiently large, the operating trajectory may remain strictly on the right-hand side of M$_1$ throughout the entire prescribed primary frequency regulation range. As a result, the inverse droop phenomenon is effectively eliminated. Nevertheless, it may lead to underutilization of the reserved power capability. As shown in Fig. \ref{fig_DeloadingHighDroop}, even when $\omega$ drops to its lower regulation limit $\omega_{min}$, the operating point B$_1$ may remain separated from M$_1$. This gap indicates that a portion of the curtailed power reserve is never utilized during the primary frequency regulation, which is overly conservative from an energy utilization perspective.

Meanwhile, the effectiveness of a high $d\%$ remains deeply dependent on wind conditions. When $v_w$ decreases from $v_{w1}$ to $v_{w2}$, the stable operating region shrinks accordingly. The operating point can still cross the compressed MPPT point M$_2$ and transition to the under-speed operating point B$_2$ under $\omega_{min}$. Under such circumstances, the inverse droop behavior reappears despite the large deloading level, as shown by the trajectory between m$_2$ and b$_2$ in Fig. \ref{fig_DeloadingHighDroop}.

Moreover, when $\omega$ increases, the rotor accelerates from A$_1$ and A$_2$ toward C$_1$ and C$_2$. Since the nominal operating point is already heavily over-speeded, the required power reduction drives the turbine toward excessively high rotor speeds and deep power curtailment compared with the low-deloading strategy for the same frequency deviation and $-p_{prc}$.

In summary, fixed-deloading strategies fail to solve the coordination problem between primary frequency regulation and GFM-WTG operation.

\section{Proposed Coordinated Control Strategy}\label{section_proposed}

\subsection{Control Objective and Design Principle}
\label{subsection_ControlObjective}

To resolve the mismatch identified in Sections \ref{section_mppt} and \ref{section_deloading}, we propose a coordinated control framework. Rather than selecting a fixed operating characteristic and subsequently evaluating its frequency response, the proposed strategy starts from the desired frequency regulation requirement. Specifically, consider that the primary frequency regulation range is predefined as $[\omega_{min},~\omega_{max}]$ and the allowable power range for frequency support is specified by a deloading limit $d_{pr}\%$. The following coordination objectives are imposed.

When $\omega$ decreases from $\omega_{st}$ to $\omega_{min}$, the operating point should continuously move toward the MPPT point. At $\omega_{min}$, it should exactly reach the MPPT point. In other words, the available power reserve should be fully utilized when the maximum frequency support requirement is reached to avoid both the inverse droop phenomenon and the inefficient utilization of the available wind resource.

When $\omega$ increases from $\omega_{st}$ to $\omega_{max}$, the operating point should move toward a lower power operating condition to provide the required downward regulation capability. At $\omega_{max}$, the WTG should exactly reach the predefined minimum operating power $p_{min}$ corresponding to the deloading limit $d_{pr}\%$. Consequently, the available power reduction capability is fully utilized while avoiding excessive power curtailment, unnecessarily large rotor acceleration, and nonutilization of available regulation capability.

Finally, these requirements should hold for varying $v_w$.

\subsection{Power-Tracking-Coefficient-Frequency Droop Control}
\label{subsec_PowerTrackingCoefficientFrequencyDroop}

\begin{figure}[!t]
\centering
\includegraphics[width=\columnwidth]{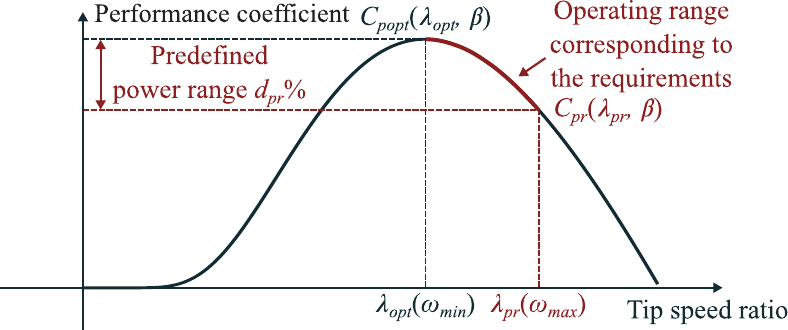}
\caption{Characteristics of performance coefficient corresponding to predefined primary frequency regulation requirements.}
\label{fig_PerformanceCoefficientPr}
\end{figure}

The discussed requirements can be directly mapped onto the $C_p$ curve of the wind turbine, as shown in Fig. \ref{fig_PerformanceCoefficientPr}. The operating point corresponding to $\omega_{min}$ is selected as the $C_{popt}$ point, characterized by $\lambda_{opt}$. Conversely, the operating point corresponding to $\omega_{max}$ is selected according to the predefined deloading limit $d_{pr}\%$ characterized by a tip-speed ratio $\lambda_{pr}$, at which $C_p$ is reduced to $C_{pr}$. Consequently, the difference between $C_{popt}$ and $C_{pr}$ directly represents the operating range allocated to primary frequency regulation. Therefore, the frequency regulation problem can be reformulated as constructing a one-to-one mapping between the frequency interval $[\omega_{min},~\omega_{max}]$ and the aerodynamic operating interval bounded by $\lambda_{opt}$ and $\lambda_{pr}$.

To realize this objective, a power-tracking-coefficient-frequency droop mechanism is proposed. In contrast to conventional GFM control, where primary frequency regulation is achieved through the additional droop term $p_{pr}$, the proposed method directly modifies the power tracking characteristics of the WTG according to the frequency deviation. Accordingly, the conventional droop regulation term is disabled in the subsequent derivation, i.e., $p_{pr} = 0$. Under this framework, the upper regulation boundary is naturally represented by the MPPT characteristic of (\ref{eq_fomegam_mppt}). Similarly, the lower boundary is represented by a predefined deloaded characteristic of $P_{st}$, given by
\begin{align}
\label{eq_fomegam_pr}
    f(\omega_m) \coloneq K_{pr}\omega_m^3
\end{align}
where $K_{pr}$ denotes the power-tracking coefficient associated with the predefined deloading limit. The value of $K_{pr}$ can be expressed as 
\begin{align}
    K_{pr} \coloneq K_{opt}\left(\lambda_{opt}/\lambda_{pr}\right)^3(1-d_{pr}\%).
\end{align}
Therefore, the original frequency regulation problem is transformed into the problem of constructing a frequency-dependent power tracking coefficient that continuously coordinates the turbine operating characteristic between $K_{opt}$ and $K_{pr}$.

Motivated by the forms of (\ref{eq_fomegam_mppt}) and (\ref{eq_fomegam_pr}), consider the following family for $P_{st}$
\begin{align}
\label{eq_fomegamk}
    f(\omega_m,~k) = k\omega_m^3
\end{align}
where $k\in[K_{pr},~K_{opt}]$ is a generalized power-tracking coefficient. For a given wind speed $v_w$ and control signal $k$, the operating point is determined by the equilibrium between $p_w$ and $P_{st}$, which can be expressed by an implicit function $F(\omega_m, k) = 0$ as:
\begin{align}
    F(\omega_m, k) &\coloneq p_w - P_{st}\notag\\
    &= 1/2\rho\pi R^2C_p(\omega_{tn}\omega_mR/v_w)v_w^3/S_n - k\omega_m^3 = 0.
\end{align}
As $C_p(\omega_m)$ is strictly decreasing on the over-speeding set $D = \{\omega_m|\omega_{tn}\omega_mR/v_w\in[\lambda_{opt}, \lambda_{pr}]\}$ from (\ref{eq_cp}) and Fig. \ref{fig_PerformanceCoefficientPr}, there is
\begin{align}
\label{eq_dwtdk}
    d\omega_m/dk &= -(\partial F/\partial k)/(\partial F/\partial\omega_m) \\&= \frac{\omega_m^3}{1/2\rho\pi R^2v_w^3/S_n(d C_p/d\omega_m)-3k\omega_m^2}<0
\end{align}
on $D$. This result proves that for any given wind speed $v_w$, an increase in $k\in[K_{pr},~K_{opt}]$ necessarily results in a monotonic decrease in the equilibrium rotor speed $\omega_m\in D$.

Meanwhile, for the operating power, there is
\begin{align}
\label{eq_dpwdk}
    \frac{dp_w}{dk} = \frac{1}{2}\frac{\rho\pi R^2v_w^3}{S_n}\frac{dC_p(\omega_m)}{dk} = \frac{1}{2}\frac{\rho\pi R^2v_w^3}{S_n}\frac{dC_p}{d\omega_m}\frac{d\omega_m}{dk}>0
\end{align}
for $k\in[K_{pr},~K_{opt}]$. Therefore, as $k$ varies continuously and monotonically from $K_{pr}$ to $K_{opt}$, the operating rotor speed $\omega_m$ and the power $p_w$ transition continuously and monotonically from the prescribed deloading point to the MPPT point. Since both derivatives of (\ref{eq_dwtdk}) and (\ref{eq_dpwdk}) are non-zero and well-defined for all $v_w>0$, the transition is proven across all valid $v_w$.

Motivated by the previous analysis, to achieve the given primary frequency regulation specifications, $k$ can be defined as a continuous function of $\omega$, $k = g(\omega)$, such that
\begin{enumerate}
    \item $\omega_1 < \omega_2\Rightarrow g(\omega_1)\ge g(\omega_2),~\forall\omega_1,~\omega_2\in[\omega_{min},~\omega_{max}]$
    \item $g(\omega_{min}) = K_{opt}$ and $g(\omega_{max}) = K_{pr}$
\end{enumerate}
where a power-tracking-coefficient-frequency droop characteristic is actually established.

\begin{figure}[!t]
\centering
\includegraphics[width=\columnwidth]{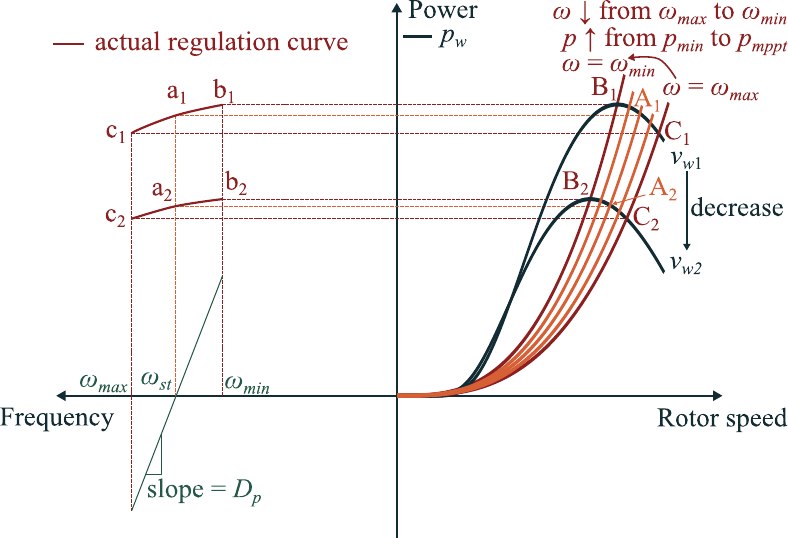}
\caption{Schematic of primary frequency regulation characteristics with the power-tracking-coefficient-frequency droop strategy.}
\label{fig_Proposed}
\end{figure}

The resulting regulation mechanism is illustrated in Fig. \ref{fig_Proposed}. At $\omega = \omega_{st}$ and $v_w = v_{w1}$, the WTG operates at an intermediate deloaded operating point A$_1$ determined by the selected function $g(\omega)$.

When $\omega$ decreases, the corresponding increase in $k$ drives the operating point toward the MPPT boundary B$_1$. Owing to the monotonic properties established above, both the captured power and the active-power output increase continuously as $\omega$ decreases. When $\omega$ decreases to $\omega_{min}$, the operating point reaches B$_1$ exactly, indicating that the entire upward power reserve has been utilized.

Conversely, when $\omega$ increases, the reduction in $k$ shifts the operating point toward the predefined deloaded boundary C$_1$ associated with $d_{pr}\%$. When $\omega$ increases to $\omega_{max}$, the operating point reaches C$_1$ exactly, thereby preventing excessive power curtailment and unnecessary rotor-speed excursions. As a result, a well-defined droop characteristic for both frequency increase and decrease will be defined within $[\omega_{min},~\omega_{max}]$, as shown between point c$_1$ and b$_1$ in Fig. \ref{fig_Proposed}.

Since the mapping $g(\omega)$ is constructed directly from the prescribed regulation requirements rather than a fixed deloading level, the resulting characteristic remains valid under different $v_w$. As illustrated in Fig. \ref{fig_Proposed}, changes from $v_{w1}$ to $v_{w2}$ merely shift the operating points to C$_2$-A$_2$-B$_2$ and the droop characteristics to c$_2$-a$_2$-b$_2$ while preserving the desired relationship between frequency and power. Thus, the strategy provides a systematic and wind-speed-robust framework for coordinating GFM control, primary frequency regulation, and aerodynamics. Notably, the power-tracking-coefficient-frequency droop also achieves the automatic power sharing among paralleled WTGs within the deloaded power range.

A simple and easy implementation of $g(\omega)$ is a linear droop characteristic, which is
\begin{align}
\label{eq_gomega}
    g(\omega) = \begin{cases}
        K_{opt} & \omega < \omega_{min}\\
        \frac{K_{opt} - K_{pr}}{\omega_{min} - \omega_{max}}(\omega-\omega_{max}) + K_{pr} & \omega_{min} \leq \omega \leq \omega_{max}\\
        K_{pr} & \omega > \omega_{max}
    \end{cases}.
\end{align}
Within the prescribed primary regulation interval $[\omega_{min},~\omega_{max}]$, $k$ varies linearly with $\omega$ to provide droop-based primary frequency regulation. Outside the interval, $k$ is saturated at its boundary values.

\subsection{Stability Analysis}

The proposed control strategy replaces the conventional droop regulation of $D_p$ with a frequency-dependent power tracking coefficient. As a result, the frequency regulation path and the associated damping characteristics change. To facilitate a comparison, a small-signal model is established.

In Fig. \ref{fig_StudiedSystem}, the output power can be calculated as
\begin{align}
\label{eq_p}
    p = VV_g/X_g\sin\delta
\end{align}
where $V$ and $V_g$ denote the filter capacitor and grid voltage magnitudes, respectively, and $X_g$ is the synchronous reactance of the grid line. Meanwhile, the power angle $\delta$ is expressed as
\begin{align}
\label{eq_delta}
    \delta = \int_0^t{\omega_n(\omega - \omega_g)}d\tau
\end{align}
where $\omega_g$ denotes the grid frequency and $\omega_n$ is the nominal angular frequency.

\begin{figure}[!t]
\centering
\includegraphics[width=\columnwidth]{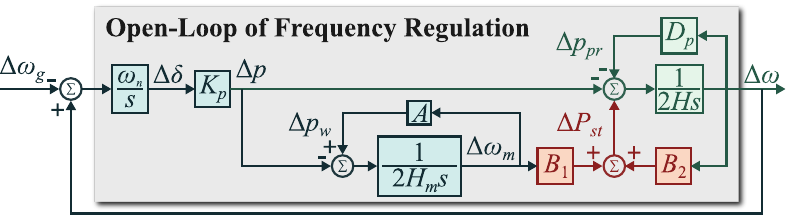}
\caption{Small-signal block diagram of frequency regulation of the system.}
\label{fig_SmallSignal}
\end{figure}

Linearizing  (\ref{eq_pw}), (\ref{eq_domegam})-(\ref{eq_ppr}), (\ref{eq_fomegamk}), (\ref{eq_p}), and (\ref{eq_delta}) yields
\begin{align}
    &\Delta p_{w} = A\Delta\omega_m \coloneqq \frac{1}{2}\frac{\rho\pi R^3v_w^2\omega_{tn}}{S_n}\left.\frac{dC_p(\lambda,~\beta)}{d\lambda}\right|_{\lambda_0}\Delta\omega_m\\
    &2H_m\Delta\dot\omega_m = \Delta p_w - \Delta p \\
    &2H\Delta\dot\omega = \Delta P_{st} - \Delta p - \Delta p_{pr}\\
    &\Delta p_{pr} = D_p\Delta\omega\\
    &\Delta P_{st} = B_1\Delta\omega_m + B_2\Delta\omega \notag\\
    & \phantom{\Delta P_{st}}\coloneq \left.\frac{\partial f}{\partial\omega_m}\right|_{(\omega_{m0},~\omega_0)}\Delta\omega_m + \left.\frac{\partial f}{\partial k}\frac{dg(\omega)}{d\omega}\right|_{(\omega_{m0},~\omega_0)}\Delta\omega \\
    & \Delta p = K_p\Delta\delta \coloneq V_0V_g/X_g\cos\delta_0\Delta\delta\\
    & \Delta\dot\delta = \omega_n(\Delta\omega - \Delta\omega_g)
\end{align}
where the resulting small-signal block diagram is shown in Fig. \ref{fig_SmallSignal}. The model provides a unified framework for comparing different strategies.

For conventional GFM applications with an ideal DC source, $P_{st}$ is treated as constant. Consequently, $\Delta P_{st} = 0$ and the corresponding open-loop transfer function becomes
\begin{align}
    G_{GFM}(s) \coloneq \frac{\Delta\omega}{\Delta\omega_g - \Delta\omega} = \frac{\omega_nK_p}{s(2Hs + D_p)}
\end{align}

For conventional deloaded GFM-WTG operation, $P_{st}$ in (\ref{eq_fomegam_deload}) depends on $\omega_m$ but remains independent of $\omega$. Therefore, $B_2 = 0$ and the resulting open-loop transfer function is
\begin{align}
    G_{deload}(s)  = \frac{\omega_nK_p(2H_ms + B_1 - A)}{s(2Hs + D_p)(2H_ms - A)}.
\end{align}
Compared with conventional GFM control, an additional dynamic mode associated with the WTG rotor is introduced. As a result, the frequency response becomes coupled with the aerodynamic operating condition.

For the proposed strategy, the primary frequency regulation action is provided through $k$ rather than $D_p$. Set $D_p = 0$ and the corresponding open-loop transfer function is
\begin{align}
\label{eq_Gproposed}
    G_{proposed}(s)  = \frac{\omega_nK_p(2H_ms + B_1 - A)}{s(2Hs - B_2)(2H_ms - A)}.
\end{align}

\begin{figure}[!t]
\centering
\includegraphics[width=\columnwidth]{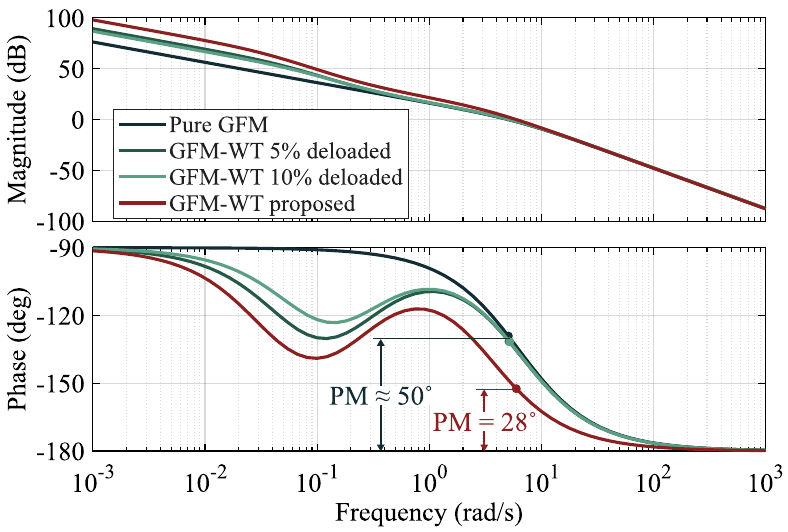}
\caption{Comparison of open-loop Bode plots of various control strategies.}
\label{fig_BodeComparison}
\end{figure}

The open-loop frequency responses of different control strategies are compared using the system parameters listed in Table \ref{tab_Parameter}. The corresponding Bode plots are shown in Fig. \ref{fig_BodeComparison}. For the conventional pure GFM control, the phase margin (PM) of the open-loop response is about $50^\circ$. Therefore, the system possesses adequate damping and satisfactory frequency regulation dynamics. When the WTG dynamics are incorporated through the conventional deloaded control strategy, this coupling mainly affects the low-frequency region and the PM remains close to that of the pure GFM case. As a result, the stability characteristics are only slightly affected. In contrast, the proposed strategy produces a pronounced phase lag around the crossover frequency as the conventional droop term $D_p$ is removed. As a result, the PM is reduced to about $28^\circ$ as shown in Fig. \ref{fig_BodeComparison}. This indicates that the proposed strategy provides weaker damping and may exhibit a more oscillatory transient response.

\begin{table}
    \centering
    \caption{GFM-PMSG-WTG System Parameters}
    \begin{tabular}{clc}
    \hline\hline
        Parameters & Description & Values\\\hline
        $S_n$ & Nominal power & 5 MW\\
        $\omega_n$ & Nominal angular frequency & 100$\pi$ rad/s\\
        $V_n$ & Nominal line-to-line RMS voltage & 690 V\\
        $\omega_{tn}$ & WT nominal rotor speed & 1.27 rad/s\\
        $\omega_{st}$ & Frequency set-point & 1 p.u.\\
        $H$ & GFM inertia constant & 5 s\\
        $D_p$ & Active power frequency droop gain & 62.5 p.u.\\
        $H_m$ & Mechanical  part inertia constant & 6.5 s\\
        $\rho$ & Air density & 1.23 kg/m$^3$\\
        $R$ & Blade radius & 63 m\\
        $\beta$ & Pitch angle & 0$^\circ$\\
        $\lambda_n$ & Tip-speed ratio at nominal frequency & 7.92\\
        $\lambda_{pr}$ & Primary regulation tip-speed ratio & 8.32\\
        $\lambda_{opt}$ & Optimal tip-speed ratio & 7\\
        $C_{popt}$ & Optimal power coefficient & 0.448\\
        $\omega_{min}$ & Minimum frequency of droop regulation & 0.992 p.u.\\
        $\omega_{max}$ & Maximum frequency of droop regulation & 1.008 p.u.\\\hline\hline
    \end{tabular}
    \label{tab_Parameter}
\end{table}

\begin{figure}[!t]
\centering
\includegraphics[width=\columnwidth]{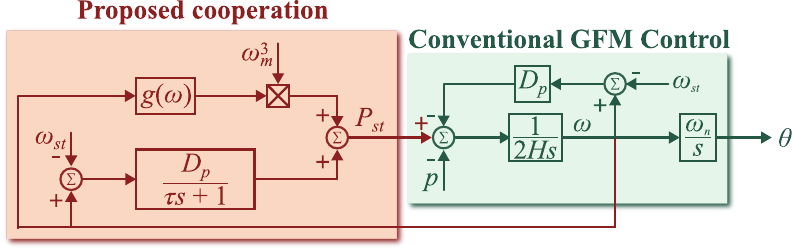}
\caption{Block diagram of the proposed coordinated strategy.}
\label{fig_ProposedBlockDiagram}
\end{figure}

\begin{figure}[!t]
\centering
\includegraphics[width=\columnwidth]{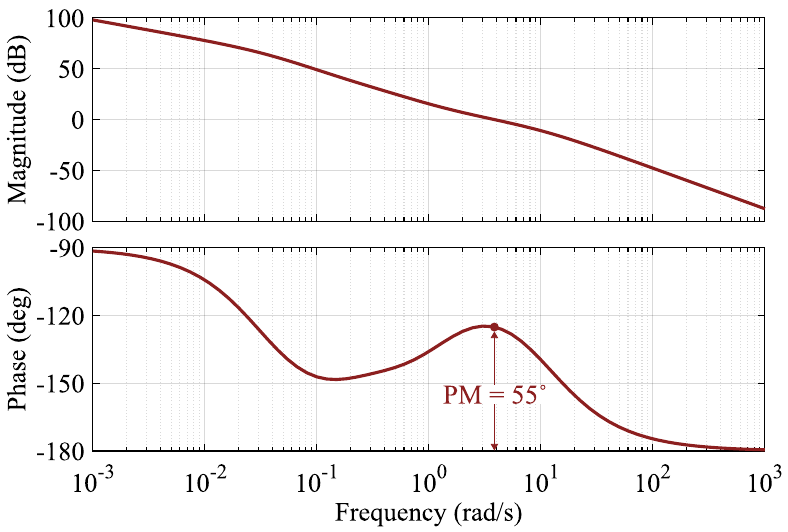}
\caption{Open-loop Bode plot of the proposed coordinated strategy.}
\label{fig_BodeProposed}
\end{figure}

To address this issue, an additional transient damping mechanism is introduced. The design objective is to recover the damping contribution of conventional droop control during transient conditions while preserving the frequency regulation characteristic established by the proposed power-tracking-coefficient-frequency droop strategy. From the perspective of the small-signal model, this objective can be achieved by augmenting the open-loop transfer function as
\begin{align}
    G_{final}(s)  = \frac{\omega_nK_p(2H_ms + B_1 - A)}{s[2Hs - B_2 + D_p\tau s/(\tau s + 1)](2H_ms - A)}
\end{align}
where $\tau$ denotes the time constant of the transient damping compensator. Up to now, the proposed strategy has removed $D_p$ and therefore modifies the original GFM control structure. To preserve compatibility with existing GFM controls, $D_p$ is retained and an additional compensation signal is introduced in $P_{st}$ to cancel it. Thus, the final $P_{st}$ is defined as
\begin{align}
    P_{st} &= g(\omega)\omega_m^3 - \frac{D_p\tau s}{\tau s + 1}(\omega - \omega_{st}) + D_p(\omega - \omega_{st})\\
    &=g(\omega)\omega_m^3 + \frac{D_p}{\tau s + 1} (\omega - \omega_{st}).
\end{align}
where the block diagram of the proposed coordinated strategy is shown in Fig. \ref{fig_ProposedBlockDiagram}. In this paper, $\tau = 0.8$ is used, which results in PM increased to $55^\circ$ as shown in Fig. \ref{fig_BodeProposed}.

In summary, the proposed coordinated strategy simultaneously achieves four objectives:
\begin{enumerate}
    \item explicit satisfaction of the prescribed primary frequency regulation limits under varying wind conditions;
    \item Automatic power sharing among paralleled units within the deloaded power range;
    \item adequate damping and stability margins;
    \item compatibility with any GFM control with droop characteristics.
\end{enumerate}

\subsection{Discussion on Flexibility of $g(\omega)$}

It should be emphasized that the choice of $g(\omega)$ is not restricted to the example in (\ref{eq_gomega}). Any $g(\omega)$ can be a candidate if the requirements explained in Section \ref{subsec_PowerTrackingCoefficientFrequencyDroop} are satisfied.

The linear characteristic used in (\ref{eq_gomega}) specifies only the two regulation boundaries. As a result, the operating point at the nominal frequency is not directly prescribed but is instead determined by both $g(\omega)$ and the aerodynamics of the WTG. To introduce an additional design degree of freedom, $g(\omega)$ can be defined using a piecewise linear characteristic as
\begin{align}
\label{eq_gomega_augement}
    g(\omega) = \begin{cases}
        K_{opt} & \omega \leq \omega_{min}\\
        \frac{K_{opt} - K_n}{\omega_{min} - \omega_{st}}(\omega - \omega_{st}) + K_n & \omega_{min} < \omega \leq \omega_{st}\\
        \frac{K_n - K_{pr}}{\omega_{st} - \omega_{max}}(\omega - \omega_{max}) + K_{pr} & \omega_{st} < \omega \leq \omega_{max}\\
        K_{pr} & \omega \geq \omega_{max}
    \end{cases}
\end{align}
where $K_n$ denotes the power tracking coefficient associated with $\omega_{st}$. It can be expressed as
\begin{align}
    K_n \coloneq K_{opt}\left(\frac{\lambda_{opt}}{\lambda_{n}}\right)^3(1-d_n\%)
\end{align}
where $\lambda_n$ and $d_n\%$ represent the prescribed deloaded tip-speed ratio and deloading level at $\omega_{st}$, respectively. The parameter $d_n\%$ determines how the available regulation range is allocated between under-frequency and over-frequency support, therefore enabling different regulation priorities to be realized without altering the prescribed boundary conditions.

\section{Case Studies}\label{section_case}

\begin{figure}[!t]
\centering
\includegraphics[width=\columnwidth]{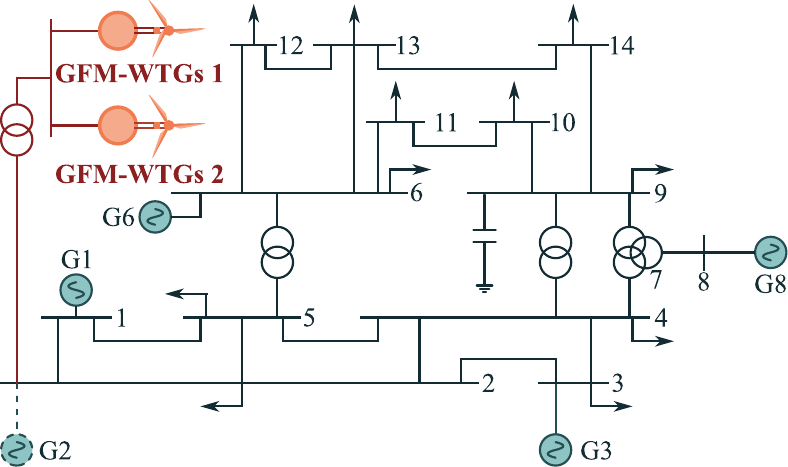}
\caption{Modified IEEE 14-bus test system.}
\label{fig_IEEE14}
\end{figure}

To evaluate the effectiveness of the proposed coordinated control strategy, case studies are conducted on a modified IEEE 14-bus test system as shown in Fig. \ref{fig_IEEE14}. The original SG connected to Bus 2 is replaced by two groups of GFM-WTGs. Each group consists of ten identical WTGs with parameters identical to Table \ref{tab_Parameter}. To investigate the influence of different wind conditions, GFM-WTGs 1 operate under a nominal wind speed of 11.4 m/s, whereas GFM-WTGs 2 operate under a lower wind speed of 8 m/s. For the proposed strategy, (\ref{eq_gomega_augement}) is used. As an example, the prescribed primary frequency regulation specification is $[\omega_{min},~\omega_{max}]=[0.992,~1.008]$ p.u. with a deloading limit $d\% = 10\%$. Meanwhile, the under-frequency and over-frequency support range is set to be identical, which means $d_n = 5\%$ for the nominal frequency. The existing deloaded GFM-PMSG-WTG control in \cite{Lyu2024} is used for comparison.

\subsection{Response to Frequency Decreasing}

\begin{figure}[!t]
\centering
\includegraphics[width=\columnwidth]{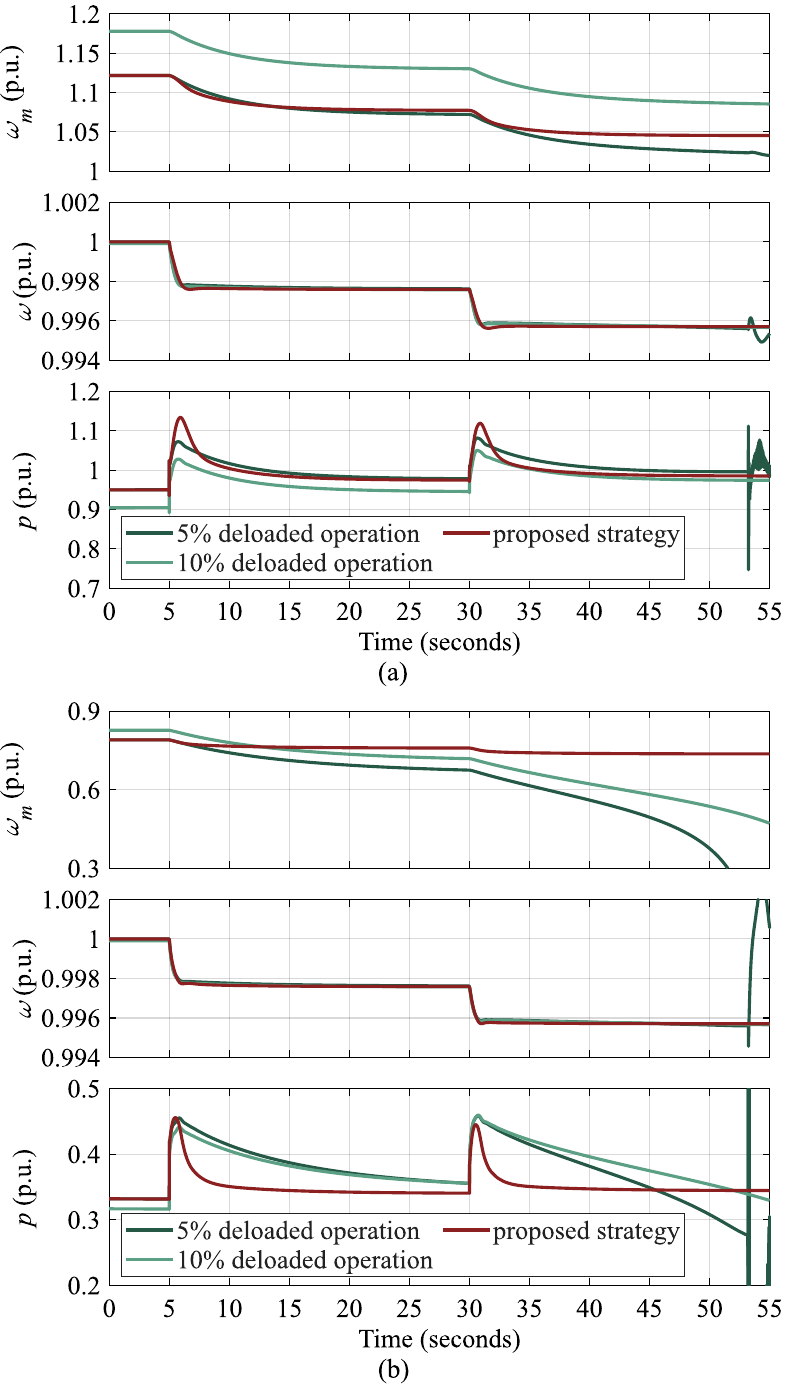}
\caption{Waveform comparison in response to load increase of (a) GFM-WT 1 and (b) GFM-WT 2.}
\label{fig_SimulationLoadIncrease}
\end{figure}

Fig. \ref{fig_SimulationLoadIncrease} compares the dynamic responses of the studied control strategies following load increases. The system initially operates close to the nominal frequency. At $t=5$ s, a 75 MW resistive load is added to Bus 2, causing $\omega$ to decrease. As expected, all GFM-WTG schemes respond to $\omega$ deviation by increasing $p$. As a result, $\omega_m$ decreases following the disturbance as shown in Fig. \ref{fig_SimulationLoadIncrease}.

For GFM-WTGs 1 with a high $v_w$ in Fig. \ref{fig_SimulationLoadIncrease}(a), the proposed strategy produces a larger increase in $p$ immediately after the disturbance. The corresponding $\omega_m$ decreases more rapidly, resulting in an improved initial frequency support capability and rate-of-change-of-frequency (RoCoF). After the initial transient period, both $\omega_m$ and $p$ of the proposed strategy settle more rapidly to a new equilibrium to maintain $\omega_m$ within a relatively narrow operating range. A more pronounced difference can be observed for GFM-WTGs 2 with a low $v_w$ shown in Fig. \ref{fig_SimulationLoadIncrease}(b). Under reduced stable operating margin (as explained in Fig. \ref{fig_DeloadingLowDroop} and \ref{fig_DeloadingHighDroop}), the proposed strategy effectively constrains $\omega_m$ excursion while maintaining frequency support. Although the constrained decrease in $\omega_m$ leads to a slightly lower frequency nadir, the influence is acceptable. In contrast, although the conventional schemes inject more active power after the initial support, helping to slightly improve the frequency nadir, this behavior is achieved at the expense of significantly larger reductions in $\omega_m$. As a result, $\omega_m$ continues to drift toward the stable operating limits.

At $t = 30$ s, an additional 75 MW load is connected to Bus 3, resulting in a second $\omega$ decline. The responses immediately following the disturbance are qualitatively similar to those observed after the first load increase. However, for GFM-WTGs 2 with the conventional deloaded strategies and a low $v_w$, $\omega_m$ continuously decreases and eventually leads to instability, which agrees with the analytical observations presented in Section \ref{section_deloading}. The stability occurs earlier for the 5\% deloaded case and is delayed, but not eliminated, even when the deloading level is increased to the prescribed limit of 10\%. Noticeably, $\omega$ settled about 0.996 p.u. is still within the prescribed limit, where the primary frequency regulation should be well performed. Another noteworthy observation is $\omega_m$ of GFM-WTGs 1 under a large $v_w$. $\omega_m$ for both of the deloading levels is still larger than 1 p.u., which means that the operating point is still located on the right of the MPPT point, remaining some power reserves, whereas the instability has been triggered by GFM-WTG 2, implying a high sensitivity to $v_w$.

In comparison, the proposed strategy maintains stable operation throughout the entire simulation. Both GFM-WTGs 1 under a high $v_w$ and GFM-WTGs 2 under a low $v_w$ remain within their admissible operating regions, demonstrating the robustness of the proposed strategy. Meanwhile, it is also noticed that following the transient response, the proposed strategy automatically adjusts the power sharing of the parallel WTGs within the prescribed power range.

\subsection{Response to Frequency Increasing}

\begin{figure}[!t]
\centering
\includegraphics[width=\columnwidth]{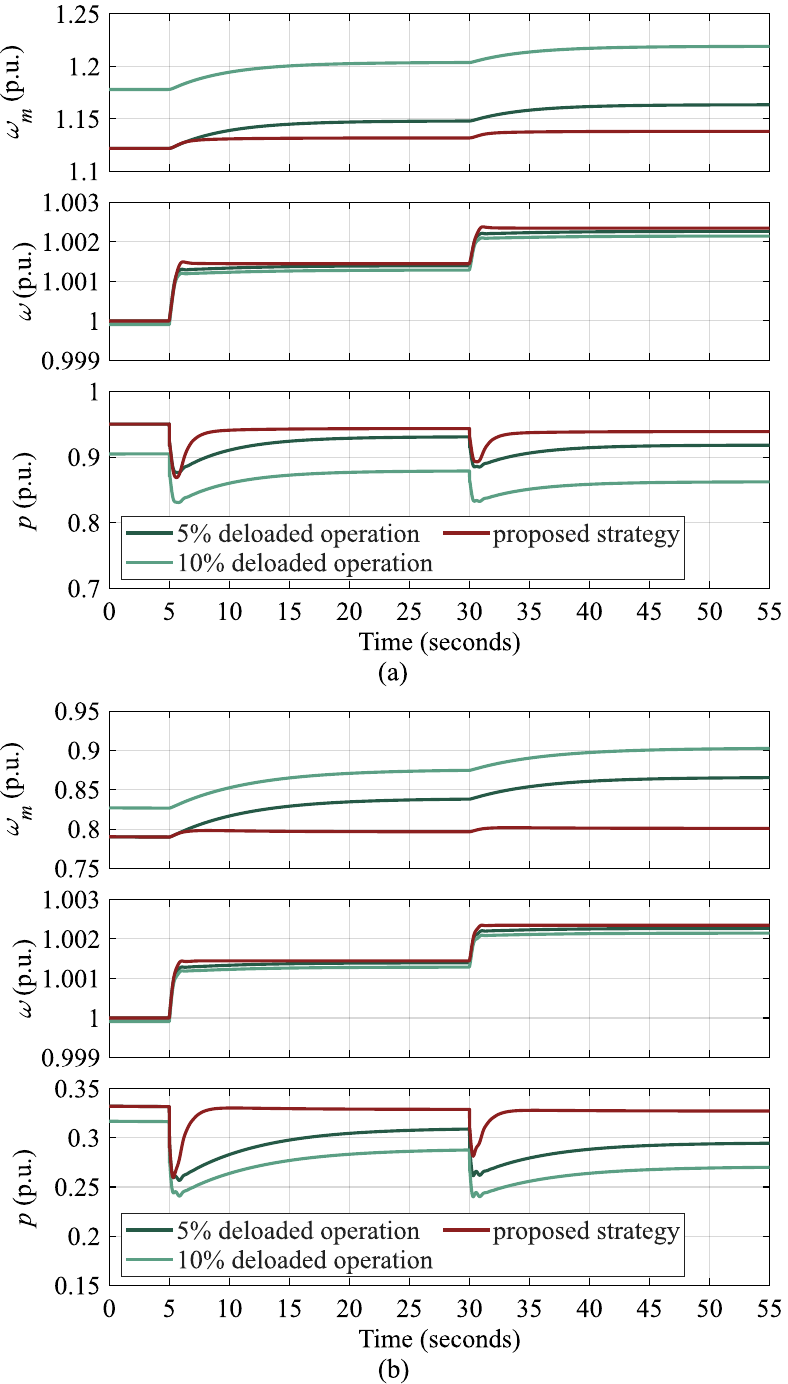}
\caption{Waveform comparison in response to load increasing of (a) GFM-WT 1 and (b) GFM-WT 2.}
\label{fig_SimulationLoadShed}
\end{figure}

Fig. \ref{fig_SimulationLoadShed} compares the responses following load decreases. At $t = 5$s, the load connected to Bus 4 (47.8 MW) is disconnected. In response, all GFM-WTG schemes reduce their output $p$ and the rotors accelerate. However, the conventional deloaded strategies exhibit substantially larger increases in $\omega_m$. This behavior is particularly evident for the 10\% deloaded case, where the nominal operating point is already located at the lower boundary of the prescribed power range. Therefore, a power reduction inevitably leads to rotor acceleration and power curtailment beyond the boundaries. A second load decrease event is applied at $t = 30$ s through the disconnection of the load connected to Bus 9 (29.5 MW). Such an event further makes the 10\% deloading scheme violate the prescribed operating constraints.

In terms of the 5\% deloaded strategy, GFM-WTGs 1 with a high $v_w$ remain within the boundary as $p$ is kept larger than 0.9 p.u. in steady-state, as shown in Fig. \ref{fig_SimulationLoadShed}(a). Nevertheless, it has approached the boundary, which indicates that only a small additional frequency increase event would be sufficient to violate the prescribed constraints. More importantly, for GFM-WTGs 2 with a low $v_w$ in Fig. \ref{fig_SimulationLoadShed}(b), after the second load decrease event, the operating point has been driven beyond the intended regulation boundary as $p$ is finally below 0.3 p.u. (the boundary for the low $v_m$ scenario of 8 m/s).

By contrast, the proposed strategy consistently maintains the operating points of both groups of GFM-WTGs, under either high or low $v_w$, within the predefined operating region throughout the simulation. Also, the proposed strategy naturally distributes the regulation effort among the paralleled GFM-WTG groups within the prescribed power range.

\section{Conclusion}\label{section_conclusion} 

This paper first investigated the fundamental mismatch between conventional GFM control and the aerodynamic characteristics of PMSG-WTGs in primary frequency regulation. It was shown that directly applying conventional GFM strategies can result in inverse droop behavior, underutilization of available power reserves, violation of operating limits, and even instability under varying wind conditions. To address these challenges, a coordinated control framework based on a power-tracking-coefficient-frequency droop mechanism was proposed. Analytic investigations demonstrated that the proposed method guarantees monotonic and well-defined frequency-power regulation characteristics within prescribed frequency and power limits while remaining adaptive to changing wind speeds. Through the introduction of a transient damping compensation mechanism, the proposed strategy also preserves adequate stability margins and can be readily integrated into existing GFM implementations with droop characteristics. Furthermore, the method naturally enables adaptive power sharing among paralleled WTGs.


%





\ifCLASSOPTIONcaptionsoff
  \newpage
\fi



\bibliographystyle{IEEEtran}
\bibliography{IEEEabrv,paper}

\begin{thebibliography}{10}
\providecommand{\url}[1]{#1}
\csname url@samestyle\endcsname
\providecommand{\newblock}{\relax}
\providecommand{\bibinfo}[2]{#2}
\providecommand{\BIBentrySTDinterwordspacing}{\spaceskip=0pt\relax}
\providecommand{\BIBentryALTinterwordstretchfactor}{4}
\providecommand{\BIBentryALTinterwordspacing}{\spaceskip=\fontdimen2\font plus
\BIBentryALTinterwordstretchfactor\fontdimen3\font minus
  \fontdimen4\font\relax}
\providecommand{\BIBforeignlanguage}[2]{{%
\expandafter\ifx\csname l@#1\endcsname\relax
\typeout{** WARNING: IEEEtran.bst: No hyphenation pattern has been}%
\typeout{** loaded for the language `#1'. Using the pattern for}%
\typeout{** the default language instead.}%
\else
\language=\csname l@#1\endcsname
\fi
#2}}
\providecommand{\BIBdecl}{\relax}
\BIBdecl

\bibitem{Chen2020}
M.~Chen, D.~Zhou, and F.~Blaabjerg, ``Impact of synchronous generator
  replacement with vsg on power system stability,'' in \emph{2020 IEEE 21st
  Work. Control Model. Power Electron. (COMPEL)}, Aalborg, Denmark, 2020, pp.
  1--7.

\bibitem{Blaabjerg2023}
F.~Blaabjerg, Y.~Yang, K.~A. Kim, and J.~Rodriguez, ``Power electronics
  technology for large-scale renewable energy generation,'' \emph{Proc. IEEE},
  vol. 111, no.~4, pp. 335--355, Apr. 2023.

\bibitem{Wang2018}
S.~Wang and K.~Tomsovic, ``A novel active power control framework for wind
  turbine generators to improve frequency response,'' \emph{IEEE Trans. Power
  Syst.}, vol.~33, no.~6, pp. 6579--6589, Nov. 2018.

\bibitem{Gonzalez2023}
A.~González-Cajigas, E.~J. Bueno, J.~Roldán-Pérez, R.~Martín-López, and
  E.~Sáiz-Marín, ``Control choices to allow the parallel operation of
  grid-forming {Type-III} wind turbines,'' \emph{IEEE Trans. Power Electron.},
  vol.~38, no.~12, pp. 15\,353--15\,364, Dec. 2023.

\bibitem{Blaabjerg2024}
F.~Blaabjerg, M.~Chen, and L.~Huang, ``Power electronics in wind generation
  systems,'' \emph{Nat. Rev. Electr. Eng.}, vol.~1, pp. 234--250, Apr. 2024.

\bibitem{Yang2018}
D.~Yang, J.~Kim, Y.~C. Kang, E.~Muljadi, N.~Zhang, J.~Hong, S.-H. Song, and
  T.~Zheng, ``Temporary frequency support of a {DFIG} for high wind power
  penetration,'' \emph{IEEE Trans. Power Syst.}, vol.~33, no.~3, pp.
  3428--3437, May 2018.

\bibitem{Bonfiglio2019}
A.~Bonfiglio, M.~Invernizzi, A.~Labella, and R.~Procopio, ``Design and
  implementation of a variable synthetic inertia controller for wind turbine
  generators,'' \emph{IEEE Trans. Power Syst.}, vol.~34, no.~1, pp. 754--764,
  Jan. 2019.

\bibitem{Zhou2023}
Y.~Zhou, D.~Zhu, X.~Zou, C.~He, J.~Hu, and Y.~Kang, ``Adaptive temporary
  frequency support for {DFIG}-based wind turbines,'' \emph{IEEE Trans. Energy
  Convers.}, vol.~38, no.~3, pp. 1937--1949, Sep. 2023.

\bibitem{Heidari2025}
M.~Heidari, L.~Ding, M.~Kheshti, X.~Zhao, and V.~Terzija, ``Adaptive inertial
  control for wind turbine generators in fast frequency response based on the
  power reduction period assessment,'' \emph{IEEE Trans. Sustain. Energy},
  vol.~16, no.~1, pp. 377--391, Jan. 2025.

\bibitem{Tang2024}
Y.~Tang, P.~Yang, Y.~Yang, Z.~Zhao, and L.~L. Lai, ``Fuzzy adaptive frequency
  support control strategy for wind turbines with improved rotor speed
  recovery,'' \emph{IEEE Trans. Sustain. Energy}, vol.~15, no.~2, pp.
  1351--1364, Apr. 2024.

\bibitem{Bao2023}
W.~Bao, L.~Ding, Y.~C. Kang, and L.~Sun, ``Closed-loop synthetic inertia
  control for wind turbine generators in association with slightly over-speeded
  deloading operation,'' \emph{IEEE Trans. Power Syst.}, vol.~38, no.~6, pp.
  5022--5032, Nov. 2023.

\bibitem{Zhao2025a}
T.~Zhao, H.~Liu, J.~Su, N.~Wang, and Z.~Luo, ``Coordinated control for
  distributed energy resources in islanded microgrids with improved frequency
  regulation capability,'' \emph{Renew. Energy}, vol. 244, May 2025.

\bibitem{Yang2024}
D.~Yang, X.~Wang, W.~Chen, G.-G. Yan, Z.~Jin, E.~Jin, and T.~Zheng, ``Adaptive
  frequency droop feedback control-based power tracking operation of a {DFIG}
  for temporary frequency regulation,'' \emph{IEEE Trans. Power Syst.},
  vol.~39, no.~2, pp. 2682--2692, Mar. 2024.

\bibitem{Zhao2025}
T.~Zhao, H.~Liu, N.~Wang, J.~Su, Z.~Luo, and S.~Ma, ``An active power control
  of {DFIG}-based wind turbine generators for frequency regulation with
  expected dynamic performance,'' \emph{Appl. Energy}, vol. 391, 2025.

\bibitem{Chen2024}
M.~Chen, D.~Zhou, A.~Tayyebi, E.~Prieto-Araujo, F.~Dörfler, and F.~Blaabjerg,
  ``On power control of grid-forming converters: Modeling, controllability, and
  full-state feedback design,'' \emph{IEEE Trans. Sustain. Energy}, vol.~15,
  no.~1, pp. 68--80, Jan. 2024.

\bibitem{Jiang2025}
S.~Jiang, Y.~Zhu, T.~Xu, X.~Wang, and G.~Konstantinou, ``Frequency domain
  inertia design of grid-forming converters,'' \emph{IEEE Trans. Power
  Electron.}, vol.~40, no.~7, pp. 8886--8898, Jul. 2025.

\bibitem{Wang2023}
G.~Wang, L.~Fu, Q.~Hu, C.~Liu, and Y.~Ma, ``Transient synchronization stability
  of grid-forming converter during grid fault considering transient switched
  operation mode,'' \emph{IEEE Trans. Sustain. Energy}, vol.~14, no.~3, pp.
  1504--1515, Jul. 2023.

\bibitem{Alghamdi2021}
B.~Alghamdi and C.~A. Cañizares, ``Frequency regulation in isolated microgrids
  through optimal droop gain and voltage control,'' \emph{IEEE Trans. Smart
  Grid}, vol.~12, no.~2, pp. 988--998, Mar. 2021.

\bibitem{Chen2022}
L.~Chen, X.~Du, B.~Hu, and F.~Blaabjerg, ``Drivetrain oscillation analysis of
  grid forming {Type-IV} wind turbine,'' \emph{IEEE Trans. Energy Convers.},
  vol.~37, no.~4, pp. 2321--2337, Dec. 2022.

\bibitem{Liu2024}
S.~Liu, H.~Wu, T.~Bosma, and X.~Wang, ``Impact of {DC}-link voltage control on
  torsional vibrations in grid-forming {PMSG} wind turbines,'' \emph{IEEE
  Trans. Energy Convers.}, vol.~39, no.~4, pp. 2631--2642, Dec. 2024.

\bibitem{Gonzalez2024}
A.~González-Cajigas, E.~J. Bueno, and J.~Roldán-Pérez, ``Compliance of
  grid-forming requirements of grid codes with a {Type III} wind turbine
  controlled as a virtual synchronous machine,'' \emph{IEEE Transactions on
  Energy Conversion}, vol.~39, no.~4, pp. 2244--2257, Dec. 2024.

\bibitem{Zhang2025}
X.~Zhang, Y.~Huang, and Y.~Fu, ``Oscillation energy transfer and integrated
  stability control of grid-forming wind turbines,'' \emph{IEEE Trans. Sustain.
  Energy}, vol.~16, no.~2, pp. 826--839, Apr. 2025.

\bibitem{Lyu2024}
X.~Lyu and D.~Groß, ``Grid forming fast frequency response for {PMSG}-based
  wind turbines,'' \emph{IEEE Trans. Sustain. Energy}, vol.~15, no.~1, pp.
  23--38, Jan. 2024.

\bibitem{Yuan2024}
X.~Yuan, Z.~Du, Y.~Li, Y.~Xu, J.~Li, D.~Yu, H.~Peng, and Z.~Xu, ``Two-stage
  coordinated control of {Type-4} wind turbine with grid-forming ability for
  active damping support,'' \emph{IEEE Trans. Energy Convers.}, vol.~39, no.~2,
  pp. 817--830, Jun. 2024.

\end{thebibliography}
%

%








\end{document}